\documentclass{article}

 \title{On the Impact of Information Technologies on Society: an
   Historical Perspective through the Game of Chess}
 \author{Fr\'ed\'eric Prost\\
         LIG, Universit\'e de Grenoble \\
         B. P. 53,  F-38041 Grenoble, France \\
        {\tt Frederic.Prost@imag.fr}}
 \date{}

\begin{document}
 
 \maketitle

 \begin{abstract}
   The game of chess as always been viewed as an iconic representation
   of intellectual prowess. Since the very beginning of computer
   science, the challenge of being able to program a computer capable
   of playing chess and beating humans has been alive and used both as
   a mark to measure hardware/software progresses and as an ongoing
   programming challenge leading to numerous discoveries. In the early
   days of computer science it was a topic for specialists.  But as
   computers were democratized, and the strength of chess engines
   began to increase, chess players started to appropriate to
   themselves these new tools. We show how these interactions between
   the world of chess and information technologies have been herald of
   broader social impacts of information technologies. The game of
   chess, and more broadly the world of chess (chess players,
   literature, computer softwares and websites dedicated to chess,
   etc.), turns out to be a surprisingly and particularly sharp
   indicator of the changes induced in our everyday life by the
   information technologies. Moreover, in the same way that chess is a
   modelization of war that captures the raw features of strategic
   thinking, chess world can be seen as small society making the study
   of the information technologies impact easier to analyze and to
   grasp.
 \end{abstract}

\section*{Chess and computer science}
Alan Turing was born when the Turk automaton was finishing its more
than a century long career of illusion\footnote{Though it can be noted
  that in 1912 Leonardo Torres y Quevedo built a real machine that
  could play King and Rook versus King endgames. It is arguably the
  first real chess playing machine built in history.}. The Turk
automaton was supposed to be a machine playing chess. Actually it was
operated by a human hidden in it (it took many years for the hoax to
be found). Last year the french chess federation suspended three
titled players. They have been convicted of cheating using chess
engines during the chess Olympiad that took place in Khanty-Mansiysk
on September 2010. In a century tables have completely turned:
nowadays it is the machine that is hidden within the player.

\smallskip
Computer science, and more broadly information technologies, have
changed the world so deeply, so quickly and so unexpectedly that it is
difficult to grasp. Economists are fond of paradoxical indexes such
as the Big-Mac index \cite{PakPol03} (illustrating the
purchasing-power parities among currencies) or the skyscraper index
\cite{LA99} (a correlation between skyscraper building and economic
crises) that underly strange and funny correlations between a priori
unrelated phenomenons. In this paper we develop such an index by
showing how the interactions between the world of chess and computer
science turn out to be particularly illuminating regarding the
societal impacts of information technologies. In the same way that
chess is a metaphor of war, we advocate that the interplay between
information technologies and the chess world can be viewed as a
metaphor of the more general issue of how information technologies and
society have interacted together. Moreover, as we will show, it has
not been exceptional for the chess world's use of information
technologies to precede mainstream uses. Thus, looking at today's
relations between the world of chess and information technologies
could be telling for the future of our digitalized era. Finally, the
world of chess is smaller than the real society. In the same way
than chess captures the essence of strategic thinking in a concise and
formalized way, the world of chess can be seen as a miniature version
of society making it much easier to grasp and analyze. 

\smallskip
The game of chess has already been used as an index of the social, and
geopolitical, situation of the world. In \cite{GK03} G. Kasparov shows
how the best chess masters (and style of play) of every epoch have
deep links with the most prominent ideas, and geopolitical
conflicts. One of the first famous chess players was Ruy Lopez, a
spanish priest of the 16th century. At the time Spain was dominating
the world and was conquering the ``new world''. Then came the
Renaissance and not surprisingly one of the best players, Domenico
Lorenzo Ponziani, was from Modena in Italy. The next century was the
one of the philosophers of the enlightment and its blind beliefs in
rationalism: the best player was of course a french,
Fran\c{c}ois-Andr\'e Philidor, and his famous saying \emph{'the pawns
  are the soul of chess'} was a clear announcement of the french
revolution. The great rivalry between France and Great-Britain during
the 18th and 19th centuries found an echo in the fights between french
and britton players: La Bourdonnais vs Mc Donnell and Saint-Amand vs
Staunton to cite but a few. The parallel between the world of chess
and that of ideas and geopolitical standings has continued until
today (with the more than famous match Fisher vs Spassky in the middle
of the cold-war). It suffices to look at the reigning world champions
to see that the world has changed: Viswanathan Anand comes from India
and the women world champion, Hou Yifan, is Chinese.

\section*{Chess as a tool to discover computer capacities}

The game of chess has been intimately related with computer science
from the early beginnings of the latter. Indeed, the founding fathers
of computer science, artificial intelligence and information theory,
respectively Alan Turing, Norbert Wiener and Claude Shannon proposed
programs and principles for chess programs in the early fifties. A
time where computers were rather product of the minds than real
objects.

\medskip
At the time the questions in computer science were very fundamental
and theoretical. The computer was a new artefact and it was not clear
at all at what it could be used for, and where its limits were,
both from a theoretical and a practical point of view. It may seem
paradoxical since a computer is a very elaborated machine, which has
not been invented or found by serendipity. Nonetheless, once created
its scope remained largely unknown. So one could wonder: how come
someone built a very elaborated machine without precisely
knowing what it would be used for? The short answer is that Turing
machines were invented as a (negative) solution to the very
fundamental question of the decidability of logic. The universal
Turing machine was a by-product of a proof regarding a theorem about
the foundations of mathematics. More precisely the question was to
find a generic method to state whether any given mathematical formula
is true or not, together with a proof of this. One could argue that
there is nothing farther from a practical perspective than this
fact.

Therefore, it is natural that the first interactions between chess and
computer science were focused on the investigation of the potentialities 
of the computer. In this perspective, the game of chess was seen as
an interesting problem in order to unveil computer capabilities. The
answers of the three founding fathers were all about the possibilities
of the computers examined along three different perspectives: 
from an algorithmical point of view, from a practical point of view and
from a philosophical/fundamental point of view. 

In \cite{Wien48} Wiener gives an algorithmic answer to the problem of
chess programming. He exposes the raw principles of chess programming
(which have not fundamentally changed until today), and shows how it
is conceptually possible to program a decision algorithm by the
combination of a minimax algorithm paired with an evaluation function
(he considered a fixed depth search).

In \cite{Shan50} Shannon gives very telling motivations on why the
game of chess is especially well suited in order to discover the
possibilities of computers. Indeed he wrote: 

\begin{quotation}
\emph{This paper is concerned with the problem of constructing a computing
routine or "program" for a modern general purpose computer which will
enable it to play chess.  Although perhaps of no practical importance,
the question is of theoretical interest, and it is hoped that a
satisfactory solution of this problem will act as a wedge in attacking
other problems of a similar nature and of greater significance.}
\end{quotation}
 
It was clear for him that the quality of play, or the strength, of the
computer were not the primary aim: he was looking for  a problem
that was not a mere computation. Indeed, he adds some very interesting
remarks that strike by their premonitory status (remember that the
paper was written in 1950 in an era where computers were barely
existing), on why the game of chess is very interesting to study:

\begin{quotation}
\emph{Machines of this general type are an extension over the ordinary use
of numerical computers in several ways. First, the entities dealt with
are not primarily numbers, but rather chess positions, circuits,
mathematical expressions, words, etc. Second, the proper procedure
involves general principles, something of the nature of judgement, and
considerable trial and error, rather than a strict, unalterable
computing process. Finally, the solutions of these problems are not
merely right or wrong but have a continuous range of "quality" from
the best down to the worst.}
\end{quotation}

In \cite{Tur53} Turing goes even deeper and unfolds the question
``Could one make a machine to play chess'' from the bare problem of
enumerating legal moves:

\begin{quotation}
  \noindent \emph{i) Could one make a machine which would obey the
    rules of chess, i.e. one which would play random legal moves, or
    which could tell one whether a given move is a legal one ?}
\end{quotation}

to deeper philosophical questions:

\begin{quotation}
  \noindent\emph{iv) Could one make a machine to play chess, and to
    improve its play, game by game, profiting from the experience?}

  \noindent \emph{To these we may add two further questions, unconnected
  with chess, which are likely to be on the tip of the reader's
  tongue.}

  \noindent \emph{v) Could one make a machine which would answer questions
  put to it, in such a way that it would not be possible to
  distinguish its answers from those of a man?} 

  \noindent \emph{vi) Could one make a machine which would have feelings
  like you and I do?}
\end{quotation}

Once again we see that at the heart of Turing's concern is the study
of the computer capabilities (from raw computations to deep
metaphysical concerns), and that chess is used as tool to discover
them.

\section*{Chess as a measure of hardware and software progress}

One of the interesting features of chess is that it can be used to
measure a rich intellectual performance (a game of chess includes
computations, spatial visualization, memory, long-term planing etc.)
in a simple and relatively unbiased way. Moreover, due to the
popularity of the game one can easily find players, or even
tournaments, in order to measure the strength of a program.

Arpad Elo designed a rating system, which bears its name, based on the
assumption that the chess performance of each player in each game is a
normally distributed random variable (see \cite{Elo78}). In doing so
the rating of a player can be objectively computed: the basic idea is
to compute the average rating of its opponents during a given
period. The performance of the player is given by his/her percentage
of wins during this period. If he/she has scored 50\%, then the
performance of the player is the average of the rating of his/her
opponents. A 75\% of wins gives a performance 200 points above the
average. The correspondance between the winning percentage and the elo
points delta is taken from a gaussian curve generally flattened at
350/400 points (it means that a player with a rating 400 points higher
than you is supposed to beat you 100\% of the times). This rating
system was first adopted by the US chess federation in 1960 and is now
in use in most of the world chess federations and the international
chess federation as well. Roughly speaking an average club player is
ranked around 1500 elo points, over 2000 elo points are national level
players. International Masters are over 2400 (around 3000 players in
the world), Great International Masters are over 2500 (around 1000
players). The top ten is above 2760 and the all time record is Garry
Kasparov's 2851 on the July 1999 and January 2000 lists.

Luke Muehlhauser has compiled the historical elo ratings of the
strongest chess engines from 1963 to 2011 in \cite{Mueh11}. It is
remarkable that starting with a rating around 1500 elo for the early
version of MacHack\footnote{Interestingly MacHack won a game vs Hubert
  Dreyfus in 1967, a professor of philosophy at MIT who was hired to
  explore the issue of artificial intelligence. He wrote an essay,
  \emph{What Computers Can't Do} \cite{Drey78} in which he exposed his
  controversial views on artificial intelligence. He also stated at
  the time that ``a ten year-old can beat the machine''. } in the mid
sixties towards Deep Rybka 3 and its estimated 3200 elo of 2011, the
slope of progress has been strikingly linear. The progress of chess
engines is hard to analyze in detail because there are so many factors
to take into account, but an intuitive explanation can be the
following: Ken Thompson (another founding father of computer science
deeply interested in computer chess) made some very interesting
experiments with Belle\footnote{Belle won every tournament and world
  championship from 1980 to 1983}, see \cite{ConTho82}. He
experimentally discovered that, on average, a single extra ply in
the search depth corresponds to 200 elo points. Thus the linear rate of
progression of chess engine can be seen as a corollary of the Moore's
law: the exponential progression of computers matches (up to some
constant) the exponential combinatorics of possible moves in a game,
and finally results in a linear progression in engines chess strength.

\medskip
It became clear to anyone in the early sixties that computers were
capable of playing decent chess, among other things. The question
slightly shifted towards the speed of improvements and the limits of
the computer strength: when would the computer be able to beat chess
experts? The society was slowly integrating the idea of computers
and, as it is the case for every major scientific breakthrough, the
reaction was a subtle blend of fears and excessiv optimism. The
perfect illustration of these contradictory feelings is, once again,
symbolically given through chess. It takes the form of a game of chess
between a computer, HAL-9000, and an astronaut in Kubrik's ``2001, A
Space Odyssey'' movie.

\section*{Chess as a pionneer of a computarized economy}

Everything changed in the eighties with the apparition of personal
computers and the democratization of electronic equipment. Computers
suddenly (in a decade give or take) ceased to be an affair of
specialists to become everybody's affair. In 1977 \emph{Fidelity
  Electronics Chess challenger 1} was the first chess computer available
for consumers. It was a computer dedicated to chess that looked like a
chess set together with a rudimental interface to input and output the
moves that looked like very much like a calculator. Here again we can
see that the use of computers by the chess world announced broader
social uses. Indeed, the more than famous ``Speak and Spell'' by Texas
Instrument (remember ``E.T. the extraterrestrial'' ?  E.T. was using a
hacked ``Speak and spell'' to call home), often presented as a
precursor of the electronic devices and toys of the eighties, was only
available from 1978.

Together with the democratization of computers, the eighties marked a
turning point in the use of computers. At first computers were used to
... compute things ! Indeed, in order to simulate complicated physical
phenomenon like weather forecasts, solving fluid mechanics equations,
computing ballistic trajectories etc. you need a lot of computational
power. Yet, together with the increasing of storage capacities another
important use of computers emerged: databases. Database management does
not require complicated computations. It is the amount, and the
structure, of information that are hard to handle by hand. It is
exactly where the computer can be useful at. Once again the chess
world understood very quickly the advantages of computers to handle a
large amount of information. \emph{Chessbase GmbH} is a company that
was founded in 1985: this company proposed a chess database that was
soon adopted by chess experts (whereas at the time chess experts did
not use the chess engines that were too weak to help them in any
way). As shows the following quote of G. Kasparov from \cite{CB25},
the arrival of chess databases changed everything in the preparation
of matches for professional players:

\begin{quotation}
\emph{In January 1987 I was back to play another 'simul' against the
  Hamburg team. This time I had two days to prepare, so we dug out the
  names of all the players and checked their records in the
  computer. It was an eye-opener for me. It took about ten minutes to
  find 192 games. If I ask my trainers to find me a game, going
  through the books, it could take days. This time, armed with the
  information I needed, I beat six of the Hamburg team and drew with
  the other two. The result, 7-1, was extraordinary. They couldn't
  believe it. Because I knew their habits, I could lead them into
  traps.}
\end{quotation}

  This prefigures in a striking way the corporation's productivity
gains thanks to computers and information digitalization. What were
repetitive tasks of information administration done by hand became
quicker and automatic. Moreover, it became possible to perform those
tasks without having to deal with a third part like a secretary (the
trainers looking for specific games in the quote of G. Kasparov),
providing a more direct access to information to managers.

\section*{Chess over the network}

Another unexpected usage of computers started in the early 1990's with
the quick democratization of networks: computers became able to
communicate and build communities of people sharing common interests
across the world. The possibilities opened by this extra feature, at
first limited to the academic world, were quickly developed to play
with distant people. The Internet Chess Server opened in January 1992
and was amongst the firsts online game servers. It was not uncommon to
find several hundreds of players simultaneously playing. Due to this
success many clones appeared in several countries: german ICS, french
ICS, dutch ICS etc. It was at a time when there was not yet a web
browser (Mosaic appeared in 1993) and where the web traffic was literaly
exploding.

Almost every features of today's social networks were already present,
albeit in a primitive way, on those chess servers. It was possible to
chat, to give a short presentation of ourselves (limited to 10 ascii
lines), to define a list of friends, a list of banned people
etc. There were also special communication channels regarding the
subject of the topic (opening theory, technical issues, french
speaking channel, etc.). At the time, in 1992, I was in the first year
of study at ``Ecole Normale Supérieure de Lyon''. I remember that I
had problems to make normal people (that is people who were not studying
computer science) understand what the internet was: the easiest way for me
was to explain that it allowed me to play a live game of chess with some
unknown person located on the other side of the earth.

It is during this period that the information spread reached the speed of
liberation. The emblematic CNN television channel is often given as
the example of the fact that the world was becoming a small village in
which everyone was going to know everything almost
instantly. Interestingly the same phenomenon occurred in the chess
world, though in a more premonitory form, notably through
M. Crowther's web site : ``\emph{The Week in Chess}'' \cite{TWIC}
(often called TWIC). The web site was collecting every week, with the
help of volunteers, all results and game scores of chess games played
in tournament through the whole world. If CNN was a world-class
broadcast news channel, it was still based on the old paradigm of
top-down approach to information. What is striking in the example of
TWIC is that it announces what had later been pompously called web
2.0 : a society more horizontal in which information does not come
from the authorities to the people, but where information are
gathered by individuals. TWIC was quickly adopted by professional
players who were feeding their databases with the latest games every
week (a phenomenon which gradually announces the digital convergence
of the following decades). It also marks the beginning of the end of
the age of obscurity (others would say the beginning of the era of
massive surveillance). Suddenly every tournament chess game became
available and known instantly around the world.

Even an individual game like chess gave rise to collective
intelligence emergence thanks to networks. In 1999 G. Kasparov played
a game in which he challenged the whole world over the internet
\cite{GK00}. Actually G. Kasparov was facing a team experts that gave
several moves among which internet individuals had to vote: the most
voted move was played. The game had a phenomenal success and it is
estimated that more than 50,000 individuals from more than 75
countries participated. From every standard the game was of a very
high quality. This game was a peculiar precursor of massively open
source projects (i.e. projects not limited to professionals computer
experts) in which a loose collection of individuals are gathered
together in order to achieve a very elaborate product. Today this is
best illustrated through the success of Wikipedia that was launched in
2001.

\section*{Digital convergence and pervasive computing}

On the front of computer chess it gradually became clear that one day
computers were going to be better player than the best humans. The
only remaining question was when such a step would be done. In 1990
A. Karpov, vice world champion at the time, lost a game in a
simultaneous event against Mephisto (a descendant of chess
challenger), and in 1992 G. Kasparov won a match of blitz games (five
minutes for the whole game) vs Fritz 2, winning 6 games, tying one but
losing 4 games. It was the first time that a computer chess program
won a game vs the world champion at speed chess. The first victory of
a chess computer in standard tournament conditions was the one of Deep
Blue vs G. Kasparov in 1996 (though Kasparov won the match
4-2). Finally, G. Kasparov lost a match vs Deep Blue in may 1997
though the history of matches was not finished: in 2002 Kramnik drew a
match vs Deep Fritz winning and losing two games.

By the beginning of the 2000s the computers, because of their 
strength and constant availability, started to be used as a sparring
partners by chess experts. Chess players tested opening ideas and were
starting to use computers to analyze their own games (looking for
tactical blunders, missed defenses etc.). It was becoming possible
for everyone to have an expert at home helping them to progress. It is
not unlike what happened in a lot of domains like music, picture and
movie editing etc. in which a lot of what was limited to professionals
became consumer grade.

But the most striking feature of the technological evolution was the
slow but steady integration of different softwares and information
sources together. It was very natural in the world of chess: to pair a
chess database together with a chess engine (that can give you some
advice on the position you are looking at) was an idea present from
the start of chess databases. As we saw, TWIC was a first way to feed
the database with a constant flow of information, by collecting every
tournament games each week. More and more tournaments were starting to
broadcast live the games: the chessboards are equipped with magnets and
moves are directly transmitted through the internet to chess servers
like ICS. On those servers you have hundreds of spectators, and chess
engines, giving their evaluation of the position, commenting the
moves. The databases started to be fed live and the computer-aided
analysis of the game also became instantaneous while few years back
you had to write a book and to analyze the game with the help of
Masters to obtain similar results.

The convergence took a step to the next level with the conjunction of
smartphones (or ultra portable computers) with pervasive internet
access. It is nowadays possible to connect to huge chess databases
through the internet almost everywhere and to have world class level
software running on your smartphone at the same time. The unexpected
result of this combination is the appearance of numerous cheating
controversies in chess tournaments at every level: from the world
championship, and the unglamorous so called ``toiletgate'' or
``Bathroom controversy'' of the 2006 chess world championship between
V. Kramnik and V. Topalov\footnote{V. Topalov's team emited suspicions
  about the fact that V. Kramnik was spending long period of times
  during in the bathrooms during the games, and that it was the only
  place not under video and radio surveillance. This controversy led
  to Kramnik forfeit in game number 5.}, to local tournaments in which
random players are seldom caught using, or suspected of having used,
electronic equipment to cheat. Perhaps the more striking and elaborate
case of cheating is the one mentioned in the introduction. Three
titled players of the french team have been convicted of cheating
during the 2010 chess olympiads. This case was remarkable in how it
illustrates the new capacities of information technologies. In a
nutshell the fraud was built like this: games were broadcasted live
from Khanty-Mansiysk, Siberia, on the internet. In France, a master
was analyzing the moves with a strong chess engine and a big chess
database. Once an interesting move or variation was found he texted
through a cell phone the move to a third player (actually the french
team coach) who would indicate the move to be played through coded
gestures. You have it all: network, databases, strong artificial
intelligence, pervasive communications. In \cite{ROG11} economist
K. Rogoff starting from this specific affair goes as far as seeing in
this the premises of a radical shift in our economy. Consider this
quotation:

\begin{quotation}
   \emph{As skilled labor becomes increasingly expensive relative to
     unskilled labor, firms and businesses have a greater incentive to
     find ways to ``cheat'' by using substitutes for high-price
     inputs.}
\end{quotation}
What is called cheating in chess translates into productivity gains in
business. K. Rogoff argue that it could be the case that a lot of
decisions that were taken by humans, and previously thought to be only
manageable by humans, could actually be automated.

\section*{The digitalized era}

   Today, the links between the world of chess and the information
technologies are more intricate than ever. In an interview for Time
 Magazine \cite{HAR11}, M. Carlsen, current number one on the chess
 rating list, said that he was not certain whether he has an actual
 chess board at his home : \emph{"I might have one somewhere. I am
  not sure"}. It gives a startling illustration of the degree of
 virtualization reached in our society. 

  One can see direct influences of the information technologies on the
game of chess. Overall chess players are tougher today, they have a
more pragmatic approach and better defensive skills than before. This
is largely due to the resilience of chess engines in difficult
positions: computers have influenced the style of play of the new
generations. Indeed a large amount of children have made their first
step as chess players vs chess machines. Another point to notice is
that it seems that today's players have a broader chess culture: thanks
to databases, it is possible to browse through thousands of games very
easily. Because of this players tend to change their openings more
often instead of repeating the same schemas in order to get over the
opponents preparation. P. Svidler (six times champion of Russia), went
as far as saying (admittedly he was half serious) that \emph{"the
  future belongs to 1. g3"}, a completely offbeat way of opening the
games just to avoid any kind of preparation and play chess.  This is
the bright side, there is a darker one.

If the chess culture is broader it is also much lighter than
before. Typically in todays tournament the average player goes to the
internet to get the pairing of the next round of his tournament. Then
he looks for the games of its future opponent in his database and
quickly spots what are the openings played by his opponent and where
are his weaknesses. Then starts the so called preparation of the game:
roughly it consists in the very quick visualization of 5 to 10 model
games (helped with a chess engine to find whether or not the opponent
make typical mistakes) just before the round.

In the preparation process the player completely relies on the machine
and its judgments. This can lead to disastrous results even at the
highest levels. In the 2004 chess world match between V. Kramnik and
P. Leko, V.Kramnik lost the 8th game without having played a single
move of his own. He blindly trusted an opening preparation (partially
based on computer evaluation) that appeared to be flawed. As a chess
trainer for kids I can testify that the faith in the machine sayings
is somewhat terrifying. Very young children can tell you that this
move is better than this one because the computer says it evaluates
the position as +0.26 for white in this variation instead of +0.17. I
am afraid that this blind faith in machines and lack of critical
spirit will generalize in our society. An interesting question is how
the tension will be resolved between this blind faith in what the
machine says vs the more open mind that machines helps to create.
Indeed, thanks to the databases and the heartless machine's evaluation, 
what would have been called ugly/crazy moves or ideas are tried. From
this point of view computer chess has been an eye-opener: chess player
become more pragmatic and less dogmatic. 

Another important evolution of chess is the ever shortening of time
controls, especially regarding online chess. When I started to play
chess online the basic time control was 2 minutes with 12 seconds
added after each move (say 8 to 10 minutes per side for the whole game
since on average a game lasts 40 moves). Nowadays it is almost
impossible to find someone to play at such a slow pace. The average
blitz game is 3 minutes per side. There is even a new time category
(standard time controls were divided along three categories: classical
chess with 2 hours for the first 40 moves, rapid chess with 20 minutes
for the whole game and blitz which was traditionally 5 minutes per
side for the whole game) called bullet or lightning for games with
less than 2 minutes per side for the whole game. Needless to say that
if it allows greater quantity, the quality of the play is severly
harmed. From a thinking and meditating game chess has become a game of
interactions and reflexes thanks to computers. Because it is so much
easier to play quickly with a mouse than with actual pieces and
clocks. The field of human-computer interactions has also its word to
say in this race: many chess server interfaces have the ``premoves''
feature. That is to allow the player to actually program his move even
before his opponent has made its own move. This is another warning for
our digitalized society: to move in advance without having actually
waited for the move of the opponent is a wonderful metaphor of a
twitter driven society.

\end{document}